\documentclass[twocolumn,floatfix,showpacs,eqsecnum,prb,superscriptaddress]{revtex4}
\usepackage{graphicx}
\usepackage{amsmath}
\usepackage{amssymb}
\usepackage{bm}

\begin{document}

\title{Scattering theory of topological invariants in nodal superconductors}
\author{J. P. Dahlhaus}
\affiliation{Instituut-Lorentz, Universiteit Leiden, P.O. Box 9506, 2300 RA Leiden, The Netherlands}
\author{M. Gibertini}
\affiliation{NEST, Istituto Nanoscienze-CNR and Scuola Normale Superiore, 56126 Pisa, Italy}
\author{C. W. J. Beenakker}
\affiliation{Instituut-Lorentz, Universiteit Leiden, P.O. Box 9506, 2300 RA Leiden, The Netherlands}
\date{August 2012}
\begin{abstract}
Time-reversal invariant superconductors having nodes of vanishing excitation gap support zero-energy boundary states with topological protection. Existing expressions for the topological invariant are given in terms of the Hamiltonian of an infinite system. We give an alternative formulation in terms of the Andreev reflection matrix of a normal-metal--superconductor interface. This allows to relate the topological invariant to the angle-resolved Andreev conductance, also when the boundary state in the superconductor has merged with the continuum of states in the normal metal. A variety of symmetry classes is obtained, depending on additional unitary symmetries of the reflection matrix. We derive conditions for the quantization of the conductance in each symmetry class and test these on a model for a 2D or 3D superconductor with spin-singlet and spin-triplet pairing, mixed by Rashba spin-orbit interaction. 
\end{abstract}
\maketitle 

\section{Introduction}
\label{intro}

The topological classification of superconductors relies on the existence of an excitation gap in the bulk of the material, that prevents transitions between topologically distinct phases.\cite{Has10,Qi11} The gap of a topological superconductor closes only at the boundary, where propagating states with a linear dispersion appear. The protected boundary states are counted by a topological invariant ${\cal Q}$, expressed either in terms of the Hamiltonian of an infinite system\cite{Ryu10} or in terms of the scattering matrix for Andreev reflection from the boundary with a normal metal.\cite{Ful12}

Nodal superconductors with time-reversal symmetry also have boundary states, forming flat bands in the middle of the bulk gap.\cite{Kas00} The same topological considerations do not apply because the gap vanishes in the bulk for certain momenta $\bm{k}$ on the Fermi surface (nodal points). Examples include the cuprate superconductors (gap $\propto k_x k_y$),\cite{Tsu00} and a variety of superconductors without inversion symmetry.\cite{Bau12} Nodal superconductors may also appear as an intermediate phase in the transition from a topological superconductor to a trivial one.\cite{Ber10,Sch10}

A topological invariant can still be constructed in a nodal superconductor for a translationally invariant boundary,\cite{Sat11,Sch11} conserving the parallel momentum $\bm{k}_{\parallel}$. The value of ${\cal Q}(\bm{k}_{\parallel})$ can only change if $\bm{k}_{\parallel}$ crosses a nodal point. This topological invariant again counts the boundary states, which are now non-propagating dispersionless states (pinned to $E=0$ for a range of $\bm{k}_{\parallel}$).

In Refs.\ \onlinecite{Sat11,Sch11} the topological invariant ${\cal Q}(\bm{k}_{\parallel})$ of a nodal superconductor takes the form of a winding number, calculated from the Hamiltonian of a translationally invariant infinite system. Here we present an alternative scattering formulation, which expresses ${\cal Q}(\bm{k}_{\parallel})$ as a trace of the Andreev reflection matrix. Since the conductance of a normal-metal--superconductor (NS) interface is expressed in terms of the same Andreev reflection matrix, this alternative formulation allows for a direct connection between the topological invariant and a transport property.

If the NS interface contains a tunnel barrier, the angle-resolved conductance $G(\bm{k}_{\parallel})$ measures the density of states and directly probes the flat surface bands as a zero-bias peak.\cite{Esc10} For a transparent interface the boundary states in the superconductor merge with the continuum in the metal, resulting in a featureless density of states, but the zero-bias peak remains.\cite{Sen01} Here we relate the height of this zero-bias peak to the value of the topological invariant. While in general this relation takes the form of an inequality, a quantized conductance,
\begin{equation}
G(\bm{k}_{\parallel})=|{\cal Q}(\bm{k}_{\parallel})|\times 2e^{2}/h,
\end{equation}
may result under certain conditions which we identify.

The outline of this paper is as follows. In the next section we formulate the scattering problem and construct the topological invariant from the Andreev reflection matrix. We make contact in Sec.\ \ref{topboundary} with the Hamiltonian formulation, by closing the system and showing that we recover the number of flat bands at the boundary. We then return to the open system and in Sec.\ \ref{GQrelation} relate the angle-resolved zero-bias conductance to the topological invariant. So far we only assumed the basic symmetries of time-reversal and charge-conjugation. The effects of additional unitary symmetries are considered in Sec.\ \ref{reflectionsym}. We apply the general theory to a model of a two-dimensional (2D) nodal superconductor in Secs.\ \ref{RashbaSC} and \ref{disorder}, including also the effects of disorder. Effects that are specific to 3D are discussed in Sec.\ \ref{sec:3D}. We conclude in Sec.\ \ref{conclusion}.

\section{Topological invariant for Andreev reflection}
\label{symmetry}

\begin{figure}[tb] 
\centerline{\includegraphics[width=0.6\linewidth]{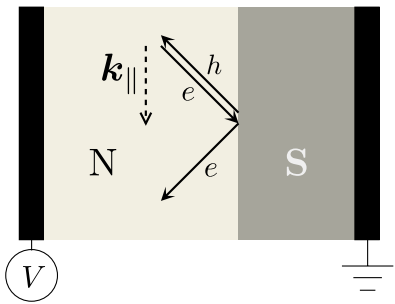}}
\caption{Interface between a superconductor (S) and a normal metal (N). The reflection matrix $r(\bm{k}_{\parallel})$ relates the amplitudes of the incident and reflected waves (arrows, both normal reflection and Andreev reflection are indicated). The conductance of the NS interface is measured by applying a voltage difference $V$ between the normal metal and the grounded superconductor.
}
\label{layout}
\end{figure}

\subsection{Chiral symmetry}
\label{chiral}

We study the Andreev reflection of electrons and holes at the Fermi level from a planar interface between a normal metal (N) and a superconductor (S). (See Fig.\ \ref{layout}.) The component $\bm{k}_{\parallel}$ along the interface of the momentum $\bm{k}$ is conserved, so we can consider each $\bm{k}_{\parallel}$ separately and work with a one-dimensional (1D) reflection matrix $r(\bm{k}_{\parallel})$. For $\bm{k}$ not in a nodal direction (nonzero excitation gap) this is a unitary matrix,
\begin{equation}
r(\bm{k}_{\parallel})r^{\dagger}(\bm{k}_{\parallel})=1.\label{unitarity}
\end{equation}

The dimension of the reflection matrix is $4\times 4$, with basis states $(\psi_{e\uparrow},\psi_{e\downarrow},\psi_{h\uparrow},\psi_{h\downarrow})$ labeled by the spin $\uparrow,\downarrow$ and the electron-hole $e,h$ degrees of freedom. The $e,h$ grading produces four $2\times 2$ submatrices,
\begin{equation}
r(\bm{k}_{\parallel})=\begin{pmatrix}
r_{ee}(\bm{k}_{\parallel})&r_{eh}(\bm{k}_{\parallel})\\
r_{he}(\bm{k}_{\parallel})&r_{hh}(\bm{k}_{\parallel})
\end{pmatrix}.\label{rehhe}
\end{equation}
Normal reflection (from electron to electron or from hole to hole) is described by $r_{ee}$ and $r_{hh}$, while $r_{he}$ and $r_{eh}$ describe Andreev reflection (from electron to hole or the other way around).

The two fundamental symmetries that we impose are time-reversal and charge-conjugation symmetry. Time-reversal symmetry requires
\begin{equation}
r(\bm{k}_{\parallel})=\sigma_{y}r^{\rm T}(-\bm{k}_{\parallel})\sigma_{y},\label{Tsymmetry}
\end{equation}
while charge-conjugation symmetry at the Fermi level requires
\begin{equation}
r(\bm{k}_{\parallel})=\tau_{x}r^{\ast}(-\bm{k}_{\parallel})\tau_{x}.\label{PHsymmetry}
\end{equation}
The Pauli matrices $\sigma_{i}$ and $\tau_{i}$ act on, respectively, the spin and electron-hole degrees of freedom. (For later use we denote the $2\times 2$ unit matrices by $\sigma_{0}$ and $\tau_{0}$.)

Taken together, Eqs.\ \eqref{Tsymmetry} and \eqref{PHsymmetry} represent the chiral symmetry relation
\begin{equation}
r(\bm{k}_{\parallel})=(\sigma_{y}\otimes\tau_{x})r^{\dagger}(\bm{k}_{\parallel})(\sigma_{y}\otimes\tau_{x}).\label{chiralsymm}
\end{equation}
This is the 1D symmetry class AIII in the periodic table of topological phases.\cite{Ryu10}

It is convenient to represent the symmetry relations in terms of the matrix $R(\bm{k}_{\parallel})=(\sigma_{y}\otimes\tau_{x})r(\bm{k}_{\parallel})$, which is both Hermitian and unitary,
\begin{equation}
R=R^{\dagger},\;\;R^{2}=1.\label{Rrelation}
\end{equation}
The submatrices in Eq.\ \eqref{rehhe} appear in $R$ as
\begin{equation}
R(\bm{k}_{\parallel})=\begin{pmatrix}
R_{he}(\bm{k}_{\parallel})&R_{hh}(\bm{k}_{\parallel})\\
R_{ee}(\bm{k}_{\parallel})&R_{eh}(\bm{k}_{\parallel})
\end{pmatrix},\label{Rehhe}
\end{equation}
where $R_{pq}=\sigma_{y}r_{pq}$. The two blocks $R_{he}$ and $R_{eh}$ are Hermitian, while $R_{ee}=R_{hh}^{\dagger}$.

\subsection{Topological invariant}
\label{secQk}

The $\mathbb{Z}$ topological invariant of 1D reflection matrices in class AIII is given by\cite{Ful11,note1}
\begin{equation}
\begin{split}
{\cal Q}(\bm{k}_{\parallel})&=\tfrac{1}{2}\,{\rm Tr}\,R(\bm{k}_{\parallel})\\
&=\tfrac{1}{2}\,{\rm Tr}\,\sigma_{y}[r_{he}(\bm{k}_{\parallel})+r_{eh}(\bm{k}_{\parallel})].
\end{split}
\label{QAIIIdef}
\end{equation}
In view of Eq.\ \eqref{Rrelation}, the $4\times 4$ matrix $R$ has eigenvalues $\pm 1$, so the value of ${\cal Q}\in\{-2,-1,0,1,2\}$. This value is $\bm{k}_{\parallel}$-independent as long as the reflection matrix remains unitary. For $\bm{k}$ in a nodal direction, the reflection matrix is sub-unitary and the topological invariant may change.

Application of Eq.\ \eqref{Tsymmetry} gives the relation
\begin{equation}
R(-\bm{k}_{\parallel})=-\tau_{x}R^{\rm T}(\bm{k}_{\parallel})\tau_{x},\label{Rminusk}
\end{equation}
which implies that
\begin{equation}
{\cal Q}(-\bm{k}_{\parallel})=-{\cal Q}(\bm{k}_{\parallel}).\label{Qminusk}
\end{equation}
If $\bm{k}_{\parallel}=0$ one necessarily has ${\cal Q}=0$. For this time-reversally invariant momentum the Pfaffian of the antisymmetric matrix $\sigma_{y}r(0)$ (equal to $\pm 1$) produces a $\mathbb{Z}_{2}$ topological invariant,\cite{Ful11,note1} characteristic of the 1D symmetry class DIII. We write this invariant in the form
\begin{equation}
{\cal Q}_{0}=1+{\rm Pf}\,\sigma_{y}r(0)\in\{0,2\},\label{Q0def}
\end{equation}
so that for ${\cal Q}_{0}$, as well as for ${\cal Q}$, the value $0$ indicates the topologically trivial phase.

\section{Topologically protected boundary states}
\label{topboundary}

The scattering formulation of topological invariants refers to an open system, without bound states. In the alternative Hamiltonian formulation, the topological invariant counts the number of dispersionless boundary states (flat bands at the Fermi level, consisting of edge states in 2D or surface states in 3D).\cite{Sat11,Sch11,Sat06,Tan09,Bry11,Sch12} To relate the two formulations, we close the system by means of an insulating barrier at the NS interface, and show that $|{\cal Q}(\bm{k}_{\parallel})|$ boundary states appear. 

The calculation closely follows Ref.\ \onlinecite{Ful11}. The number of boundary states at $\bm{k}_{\parallel}$ equals the number of independent solutions $\psi$ of
\begin{equation}
\bigl[1-r_1(\bm{k}_{\parallel})r(\bm{k}_{\parallel})\bigr]\psi=0.
\label{endstatecond}
\end{equation}
The unitary matrix $r_1$ is the reflection matrix of the barrier, approached from the side of the superconductor. We can write this equation in terms of Hermitian and unitary matrices $R_{1}=r_{1}(\sigma_{y}\otimes\tau_{x})$ and $R_{2}=(\sigma_{y}\otimes\tau_{x})r$, which we decompose as
\begin{equation}
R_i=U_i D_i U_i^\dagger,\;\;D_{i}=\begin{pmatrix}
\openone_{2+{\cal Q}_i} & 0\\
0 & -\openone_{2-{\cal Q}_i}
\end{pmatrix}.\label{RUdecomp}
\end{equation}
(The notation $\openone_{M}$ indicates the $M\times M$ unit matrix and $U_{1},U_{2}$ are unitary matrices.) Eq. (\ref{endstatecond}) takes the form
\begin{equation}
(1-D_1 U D_2 U^\dagger) \psi'=0,
\label{eq:edgestates1}
\end{equation}
with $U=U_1^\dagger U_2$ and $\psi'=U_1^\dagger \psi$.

We decompose $U$ into $N\times M$ submatrices $A_{N,M}$,
\begin{align}
U=\begin{pmatrix}
A_{2+{\cal Q}_1,2+{\cal Q}_2} & A_{2+{\cal Q}_1,2-{\cal Q}_2}\\
A_{2-{\cal Q}_1,2+{\cal Q}_2} & A_{2-{\cal Q}_1,2-{\cal Q}_2}
\end{pmatrix}.
\end{align} 
Since
\begin{align}
U-D_1 U D_2 =2\begin{pmatrix}
0& A_{2+{\cal Q}_1,2-{\cal Q}_2}\\
A_{2-{\cal Q}_1,2+{\cal Q}_2} & 0
\end{pmatrix},
\end{align}
we can rewrite Eq.~(\ref{eq:edgestates1}) as
\begin{align}
\begin{pmatrix}
0& A_{2+{\cal Q}_1,2-{\cal Q}_2}\\
A_{2-{\cal Q}_1,2+{\cal Q}_2} & 0
\end{pmatrix}\psi''=0,
\label{eq:edgestates2}
\end{align}
with $\psi''=U_2^\dagger\psi$. 

For any matrix $A_{N,M}$ with $N<M$ there exist at
least $M-N$ independent vectors $\bm{v}$ of rank $M$ such that
$A_{N,M}\bm{v} = 0$. Therefore Eq.~(\ref{eq:edgestates2}) has at least 
$|{\cal Q}_1+{\cal Q}_2|$ independent solutions. These are the topologically protected boundary states.

Because the insulating barrier is topologically trivial, ${\cal Q}_{1}=0$, while ${\cal Q}_2={\cal Q}$ is the topological invariant of the superconductor, so it all works out as expected: The topological invariant of the open system counts the number of boundary states that would appear if we would close it.

Both values ${\cal Q}$ and $-{\cal Q}$ of the topological invariant produce the same number ${\cal N}=|{\cal Q}|$ of boundary states if the superconductor is terminated by a topologically trivial barrier (an insulator or vacuum). The sign of the topological invariant matters if we consider the interface between two topologically nontrivial superconductors $1,2$. The combined number of boundary states ${\cal N}_{\rm total}=|{\cal Q}_{1}+{\cal Q}_{2}|=|{\cal N}_{1}\pm {\cal N}_{2}|$ is the sum or difference of the individual numbers depending on whether the topological invariants have the same or opposite sign.

\section{Relation between conductance and topological invariant}
\label{GQrelation}

By considering an open system when formulating the topological invariant, we can make direct contact to transport properties. The angle-resolved zero-bias conductance of the NS interface is given by
\begin{equation}
G(\bm{k}_{\parallel})=G_{0}\,{\rm Tr}\,r_{he}^{\vphantom{\dagger}}(\bm{k}_{\parallel})r_{he}^{\dagger}(\bm{k}_{\parallel}),\label{Gdef}
\end{equation}
with $G_{0}=2e^{2}/h$ the Andreev conductance quantum. We wish to relate this transport property to the topological invariant \eqref{QAIIIdef}.

For that purpose it is convenient to work with the matrices $R_{he}=\sigma_{y}r_{he}$ and $R_{eh}=\sigma_{y}r_{eh}$, since these are Hermitian (unlike the $r_{he}$ and $r_{eh}$ themselves). For brevity we omit the label $\bm{k}_{\parallel}$. The squares $R_{he}^{2}$ and $R_{eh}^{2}$ have the same set of Andreev reflection eigenvalues $\rho_{n}\in[0,1]$, which are also the eigenvalues of $r_{he}^{\vphantom{\dagger}}r_{he}^{\dagger}$.  

On the one hand we have the conductance
\begin{equation}
G/G_{0}={\rm Tr}\,R_{he}^{2}={\rm Tr}\,R_{eh}^{2},\label{GReh}
\end{equation}
and on the other hand the topological invariant
\begin{equation}
{\cal Q}=\tfrac{1}{2}\,{\rm Tr}\,(R_{he}+R_{eh}).\label{QReh}
\end{equation}
In App.\ \ref{unitrhonproof} we prove that at least $|{\cal Q}|$ of the $\rho_{n}$'s are equal to unity. This immediately implies the inequality
\begin{equation}
G/G_{0}\geq |{\cal Q}|.\label{GQinequality}
\end{equation}
For $\bm{k}_{\parallel}=0$ we have, additionally,
\begin{equation}
G/G_{0}\geq {\cal Q}_{0},\;\;{\rm for}\;\;\bm{k}_{\parallel}=0.\label{GQ0inequality}
\end{equation}

In a topologically trivial system, with ${\cal Q},{\cal Q}_{0}=0$, these inequalities are ineffective, while for $|{\cal Q}|,{\cal Q}_{0}=2$ the inequalities are saturated (since $G$ cannot become larger than $2G_{0}$). Scattering events in the normal or superconducting region that conserve $\bm{k}_{\parallel}$, such as spin mixing, cannot change the conductance once it is saturated.

\section{Effects of additional unitary symmetries}
\label{reflectionsym}

\begin{table*}[htb]
\centering
\begin{tabular}{ | c || c | c | c | c | c | c | }
\hline
$a,b$   & $x,x$ or $z,x$ & $x,0$ or $z,0$ & $x,z$ or $y,x$     & $y,y$ or $0,z$ & $x,y$ or $z,y$ & $0,0$  \\ 
           & or $0,y$         & or $y,z$          & or $z,z$ or $y,0$  &                      & or $0,x$         &            \\ \hline\hline
${\cal T}_{ab}^{2}$ & $+1$ & $+1$ & $+1$ & $-1$ & $-1$  & $-1$    \\ \hline
${\cal C}_{ab}^{2}$ & $+1$ & $+1$ & $-1$ & $-1$ & $+1$ &  $+1$    \\ \hline
class                          &   BDI & BDI   &  CI   & CII   &   DIII  &   DIII   \\ \hline
${\cal Q}_{ab}$         &$0,\pm 1,\pm 2$&$0,\pm 1,\pm 2$&$0$&$0,\pm 2$&$0,2$&$0,2$ \\ \hline
$G/G_{0}$         &$\mbox{}\geq |{\cal Q}_{ab}|$&$\mbox{}= |{\cal Q}_{ab}|$&$\times$&$\mbox{}= |{\cal Q}_{ab}|$&$\times$&$\mbox{}= |{\cal Q}_{ab}|$ \\ \hline

\end{tabular}
\caption{The first row lists the spatial symmetry \eqref{rreflection}; the second and third rows give the square of the anti-unitary operators \eqref{calTsymmetry} and \eqref{calPHsymmetry}; the fourth and fifth rows show the corresponding symmetry class and the values taken by the topological invariant; finally, the last row gives the relation between conductance and invariant for a topologically \textit{nontrivial} system (so for ${\cal Q}_{ab}\neq 0$, with $\times$ indicating the absence of a relation).}
\label{tablereflect}
\end{table*}

Further unitary symmetries may enforce restrictions on both the topological invariant and the angle-resolved conductance, or even introduce new topological invariants. In the first subsection we consider spatial symmetries that invert $\bm{k}_{\parallel}\mapsto-\bm{k}_{\parallel}$, whereas in the second subsection we address symmetries that conserve $\bm{k}_{\parallel}$.

\subsection{Spatial symmetries}

We consider a spatial symmetry of the form
\begin{equation}
r(\bm{k}_{\parallel})=(\sigma_{a}\otimes\tau_{b})r(-\bm{k}_{\parallel})(\sigma_{a}\otimes\tau_{b}).\label{rreflection}
\end{equation}
Combined with time-reversal symmetry \eqref{Tsymmetry} and charge-conjugation symmetry \eqref{PHsymmetry}, this produces the two symmetry relations
\begin{align}
&r(\bm{k}_{\parallel})={\cal T}_{ab}r^{\dagger}(\bm{k}_{\parallel}){\cal T}_{ab}^{-1},\;\;
{\cal T}_{ab}=(\sigma_{a}\cdot\sigma_{y})\otimes\tau_{b}\,{\cal K},
\label{calTsymmetry}\\
&r(\bm{k}_{\parallel})={\cal C}_{ab}r(\bm{k}_{\parallel}){\cal C}_{ab}^{-1},\;\;
{\cal C}_{ab}=\sigma_{a}\otimes(\tau_{b}\cdot\tau_{x}){\cal K},
\label{calPHsymmetry}
\end{align}
where ${\cal K}$ is the operator of complex conjugation. The product of ${\cal T}_{ab}$ and ${\cal C}_{ab}$ brings us back to the chiral symmetry \eqref{chiralsymm}.

\subsubsection{Topological invariant}
\label{restrictions}

Depending on whether the anti-unitary operators ${\cal T}_{ab}$ and ${\cal C}_{ab}$ square to $+1$ or $-1$, the reflection matrix falls in one of the four Altland-Zirnbauer symmetry classes BDI, CI, CII, DIII.\cite{Alt97} The various cases are listed in Table \ref{tablereflect}. These all have a higher symmetry than the class AIII from which we started (with only chiral symmetry). The additional symmetry may restrict the topological invariant to a smaller range of values. In class DIII a new $\mathbb{Z}_{2}$ topological invariant appears, that can be nonzero even if the $\mathbb{Z}$ invariant vanishes.

We denote the modified topological invariant by ${\cal Q}_{ab}(\bm{k}_{\parallel})$. In class CI only topologically trivial systems exist,\cite{Ryu10} meaning that the spatial symmetry allows only for ${\cal Q}_{ab}=0$. For the other three symmetry classes the topological invariants are given by\cite{Ful11}
\begin{align}
&{\cal Q}_{ab}=\tfrac{1}{2}\,{\rm Tr}\,R\in\{-2,-1,0,1,2\},\;\;\mbox{for BDI},\label{QabBDI}\\
&{\cal Q}_{ab}=\tfrac{1}{2}\,{\rm Tr}\,R\in\{-2,0,2\},\;\;\mbox{for CII},\label{QabCII}\\
&{\cal Q}_{ab}=1+{\rm Pf}\,(\sigma_{a}\otimes\tau_{b})(\sigma_{y}r)\in\{0,2\},\;\;\mbox{for DIII}.\label{QabDIII}
\end{align}
The restriction to even integers in class CII (a $2\mathbb{Z}$ invariant) is a consequence of the Kramers degeneracy of the eigenvalues of the Hermitian matrix $R=(\sigma_{y}\otimes\tau_{x})r$. Symmetry class DIII has a $\mathbb{Z}_{2}$ invariant.

\subsubsection{Conductance}
\label{quantizedG}

The expressions \eqref{QabBDI} and \eqref{QabCII} for ${\cal Q}_{ab}$ in class BDI and CII are the same as the expression \eqref{QAIIIdef} for ${\cal Q}$ in class AIII, so the topological invariant still provides a lower bound on the angle-resolved conductance,
\begin{equation}
G/G_{0}\geq |{\cal Q}_{ab}|,\;\;\mbox{for BDI and CII}.\label{GQab}
\end{equation}

In symmetry class DIII the invariants ${\cal Q}_{00}$ in Eq.\ \eqref{QabDIII} and ${\cal Q}_{0}$ in Eq.\ \eqref{Q0def} also have the same expression, so the inequality \eqref{GQ0inequality} still applies,
\begin{equation}
G/G_{0}\geq {\cal Q}_{00}.\label{GQ00}
\end{equation}
No relation with the conductance exists for the other invariants in class DIII, so ${\cal Q}_{0x},{\cal Q}_{xy}$, and ${\cal Q}_{zy}$ provide no restriction on the conductance.\cite{note2}

The inequality \eqref{GQab} can be sharpened further in class BDI, so that it becomes an equality not only for $|{\cal Q}_{ab}|=2$ but also for $|{\cal Q}_{ab}|=1$.\cite{Mat12} As we show in App.\ \ref{Berideg}, this equality is enforced by the spatial symmetry \eqref{rreflection} for $(a,b)\in\{(y,z),(x,0),(z,0)\}$, so for three out of the six symmetries in class BDI.

The last row of Table \ref{tablereflect} summarizes the relation between the topological invariant and the conductance for a topologically nontrivial system (${\cal Q}_{ab}\neq 0$). It is an equality for all symmetries in class CII and for some symmetries in classes BDI and DIII.

\subsection{Symmetries that preserve $\bm{k}_{\parallel}$}
\label{extrasym}

A different type of unitary symmetry preserves parallel momentum, 
\begin{equation}
r(\bm{k}_{\parallel})=(\sigma_{a}\otimes\tau_{b})r(\bm{k}_{\parallel})(\sigma_{a}\otimes\tau_{b}).\label{rconservation}
\end{equation}
Combined with the chiral symmetry relation \eqref{chiralsymm} and unitarity of $r$, this symmetry ensures that the matrix
$\tilde{R}=(\sigma_{a}\otimes\tau_{b})R$ is a unitary matrix that squares to $\pm 1$. We can thus define a new $\mathbb{Z}$ invariant
\begin{equation}
\tilde{\cal Q}(\bm{k}_{\parallel})=\begin{cases}
\tfrac{1}{2}\,{\rm Tr}\,\tilde{R}(\bm{k}_{\parallel})&{\rm if}\;\;\tilde{R}^{2}=1,\\
\tfrac{1}{2}i\,{\rm Tr}\,\tilde{R}(\bm{k}_{\parallel})&{\rm if}\;\;\tilde{R}^{2}=-1.
\end{cases}\label{QprimeAIIIdef}
\end{equation}

In general, $\tilde{\cal Q}$ and ${\cal Q}$ are distinct, and in particular $\tilde{\cal Q}$ can be an even function of $\bm{k}_{\parallel}$. The coexistence of two distinct topological invariants is quite unusual, and as we will see, it has observable consequences in the conductance.

For $b\in\{0,z\}$ nonzero values of $\tilde{\cal Q}$ constrain the conductance in the same way that ${\cal Q}$ does in Eq. (\ref{GQinequality}). For $b\in\{x,y\}$ one has instead the constraint
\begin{equation}
G/G_0 \le 2-|\tilde{\cal Q}|,\label{GtildeQconstraint}
\end{equation}
as we show in App.\ \ref{app:condconstr3}.

\section{Application: 2D Rashba superconductor}
\label{RashbaSC}

As a first application of our general scattering theory we consider a two-dimensional superconductor with spin-singlet and spin-triplet pairing mixed by Rashba spin-orbit coupling. The topologically protected edge states for this Rashba superconductor have been studied in Refs.\ \onlinecite{Sat11,Tan10,Yad11} using the Hamiltonian formulation. We summarize those results in the next subsection, before proceeding to the scattering formulation and the calculation of the conductance.

\subsection{Hamiltonian and edge states}

\begin{figure}[tb]
\centerline{\includegraphics[width=0.8\linewidth]{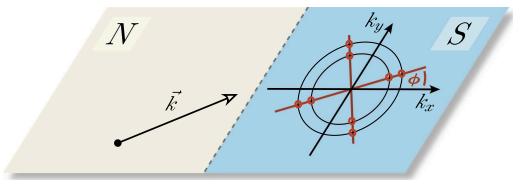}}
	\caption{Interface between a normal metal and a 2D Rashba superconductor. The Fermi surface is split into two circles, which intersect the nodal lines (red) of the superconducting pair potential in eight nodal points.
		\label{fig:dwave}}
\end{figure}

The superconductor has the Bogoliubov-de Gennes Hamiltonian
\begin{align}
H(\bm{k})=\begin{pmatrix}
\epsilon(\bm{k})+\bm{g}(\bm{k}) \cdot \bm{\sigma}&\Delta(\bm{k})\\
\Delta^\dagger(\bm{k})&-\epsilon(\bm{k})+\bm{g}(\bm{k}) \cdot \bm{\sigma}^{\ast}
\end{pmatrix},
\label{eq:Hamiltonian}
\end{align}
with free electron part $\epsilon(\bm{k})=|\bm{k}|^{2}/2m-\mu$, at Fermi energy $\mu$, and Rashba spin-orbit coupling $\bm{g}(\bm{k})=\lambda(k_y,-  k_x,0)$. We have set $\hbar=1$ and have collected the three Pauli matrices in a vector $\bm{\sigma}=(\sigma_{x},\sigma_{y},\sigma_{z})$. The Fermi surface consists of two concentric circles at momenta
\begin{equation}
k_\pm=[(m \lambda)^2+2m\mu]^{1/2}\pm m\lambda.
\label{eq:k1k2}
\end{equation}
For later use we give the spin orbit energy $E_{\rm so}=m \lambda^2$ and the spin-orbit momentum and length $k_{\rm so}=m\lambda=1/l_{\rm so}$.

The mixed singlet-triplet pair potential is given by
\begin{align}
\Delta(\bm{k})&=f(\bm{k})\left( \Delta_{\rm s}+\Delta_{\rm t} \frac{\bm{g}(\bm{k})\cdot \bm{\sigma}}{\lambda (2m\mu)^{1/2}}\right)i\sigma_y,\label{Deltafdef}\\
f(\bm{k})&=\frac{1}{2m\mu}\left[k_x k_y \cos 2\phi+\tfrac{1}{2}(k_y^2-k_x^2)\sin 2\phi\right],
\end{align}
The strength of the singlet and triplet pairing is parameterized by the energies $\Delta_{\rm s}$ and $\Delta_{\rm t}$.
The nodal lines of vanishing pair potential are oriented at an angle $\phi$ with the NS interface (see Fig.\ \ref{fig:dwave}). The intersection of the nodal lines with the Fermi surface defines 8 nodal points, in each of which ${\rm Det}\, H=0$.

The chiral symmetry
\begin{align}
H(\bm{k})=-(\sigma_{y}\otimes\tau_{x})H(\bm{\bm{k}})(\sigma_{y}\otimes\tau_{x})
\label{eq:Chiral}
\end{align}
ensures that $H$ can be brought in the off-diagonal form
\begin{equation}
{\cal U}^{\dagger}H(\bm{k}){\cal U}=\begin{pmatrix}
0 & q(\bm{k})\\
q^\dagger(\bm{k}) & 0
\end{pmatrix}.
\label{eq:transformedH}
\end{equation}
The $\mathbb{Z}$ topological invariant is then defined by the winding number\cite{Sat11}
\begin{equation}
{\cal W}(k_y)=\frac{1}{2\pi}\text{Im} \int \!dk_x\,\frac{\partial}{\partial k_x} \ln{\rm Det}\, q(k_x,k_y),
\label{eq:TIHamiltonian}
\end{equation}
for any $k_{y}$ that is not equal to the projection of one of the nodal points on the $y$-axis.

\begin{figure}[tb]
\centerline{\includegraphics[width=0.6\linewidth]{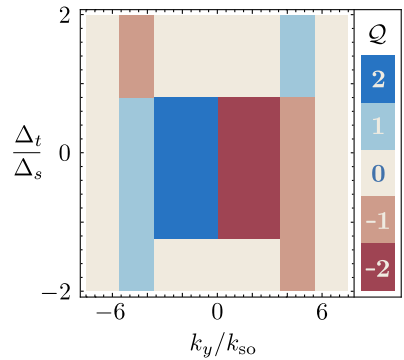}}
\caption{Topological invariant ${\cal Q}=-{\cal W}$ of the 2D Rashba superconductor ($\phi=0$, $\mu=10\,E_{\rm so}$), as a function of momentum $k_y$ along the NS interface and ratio $\Delta_{\rm t}/\Delta_{\rm s}$ of triplet and singlet pairing energies.\label{fig:phasediagram}}
\end{figure}

As analysed in Refs.\ \onlinecite{Sat11,Tan10,Yad11}, the termination of the superconductor at $x=0$ by an insulator (or by vacuum) produces $|{\cal W}(k_{y})|$ dispersionless edge states (flat bands). A simple example occurs for $\phi=0$ and $\Delta_{\rm t}=0$, corresponding to $d_{xy}$-wave spin-singlet pairing. Then
\begin{equation}
{\cal W}(k_y)=\left\{\begin{array}{cl}
2\,{\rm sign}\,(k_y)&{\rm if}\;\;|k_y|<k_-,\\
{\rm sign}\,(k_y)&{\rm if}\;\;k_-<|k_y|<k_+,\\
0&{\rm if}\;\;|k_y|>k_+,
\end{array}\right.
\label{eq:TIHamiltonian2}
\end{equation}
so there are two topologically protected edge states for $|k_y|<k_-$ and a single one for $k_-<|k_y|<k_+$.

For nonzero $\Delta_{\rm t}$ the phase boundaries \eqref{eq:TIHamiltonian2} remain unaffected in the interval
\[
-\sqrt{2m\mu}/k_- <\Delta_{\rm t}/\Delta_{\rm s}<\sqrt{2m\mu}/k_+,
\]
see Fig.\ \ref{fig:phasediagram}. To contrast the spin-singlet and spin-triplet dominated regimes, we will in what follows focus on the two limits $\Delta_{\rm t}\rightarrow 0$ and $\Delta_{\rm s}\rightarrow 0$.

\subsection{Reflection matrix and conductance}

\begin{figure}[tb]
\centerline{\includegraphics[width=1\linewidth]{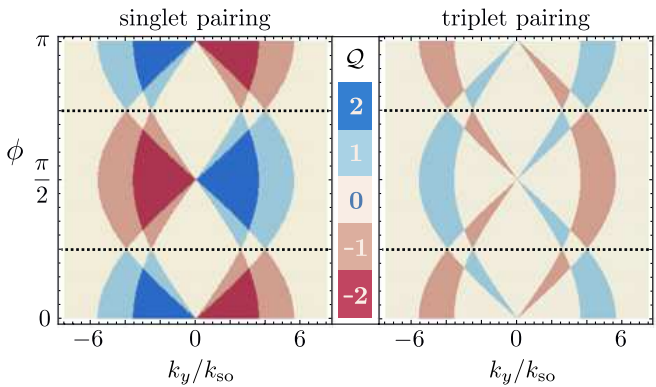}}
	\caption{Topological invariant ${\cal Q}$ of the reflection matrix from the 2D Rashba superconductor, as a function of momentum $k_y$ along the NS interface and angle $\phi$ between the interface and the nodal line. The left panel shows results for spin-singlet pairing ($\Delta_{\rm s}=E_{\rm so}$, $\Delta_{\rm t}=0$) and the right panel for spin-triplet pairing ($\Delta_{\rm t}=E_{\rm so}$, $\Delta_{\rm s}=0$). In both panels $\mu=10\, E_{\rm so}$ and $\mu_{\rm N}=30\, E_{\rm so}$. The dotted lines indicate a topologically trivial system in class CI, as a consequence of the spatial symmetry \eqref{eq:spatialsym1}.
\label{fig:TopInv}}
\end{figure}

If the superconductor is not terminated at $x=0$ but connected to a normal metal, the edge states hybridize with the continuum of the metallic bands. The topological signature then shows up in the conductance rather than in the density of states. To reveal these signatures we construct the reflection matrix of the NS interface and calculate both the topological invariant \eqref{QAIIIdef} and the angle-resolved conductance \eqref{Gdef}. 

We used either an analytical method of calculation (matching wave functions at the NS interface), or a numerical method (discretizing the Hamiltonian \eqref{eq:Hamiltonian} on a square lattice and calculating the Green function). We made sure that the lattice constant was sufficiently small that the two methods gave equivalent results. In the normal metal we set both the pair potential and the spin-orbit coupling to zero, so that there is a single Fermi circle with momentum $k_{\rm N}=(2m\mu_{\rm N})^{1/2}$. Because of a potential step at the NS interface, the chemical potential $\mu_{\rm N}$ in the normal metal ($x<0$) can differ from the value $\mu$ in the superconductor ($x>0$).

\begin{figure}[tb]
\centerline{\includegraphics[width=0.7\linewidth]{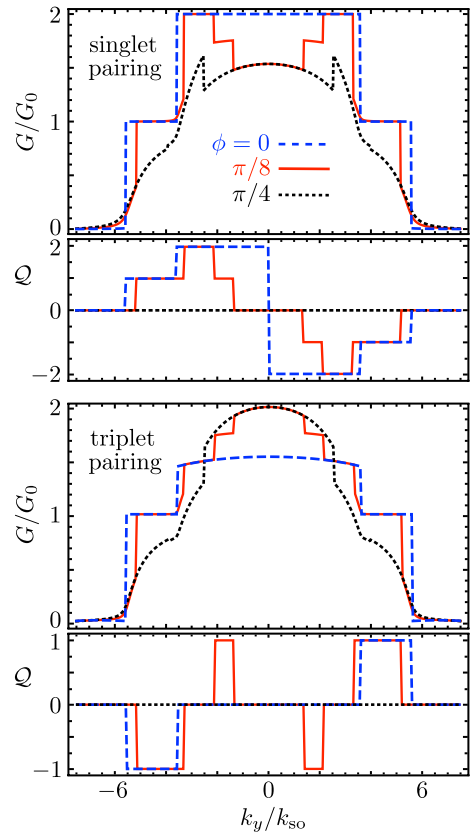}}
	\caption{Electrical conductance and $\mathbb{Z}$ topological invariant for three of the angles $\phi$ from Fig.\ \ref{fig:TopInv}. A  nonzero $\mathbb{Z}_{2}$ invariant appears in the spin-triplet case: ${\cal Q}_0=2$ for $k_y=0$, $\phi\neq 0$.
	\label{fig:Cond}}
\end{figure}

Results are collected in Figs.\ \ref{fig:TopInv} and \ref{fig:Cond}. As a first check, we note that for $\phi=0$, $\Delta_{\rm t}=0$, we recover Eq.\ \eqref{eq:TIHamiltonian2} --- up to an irrelevant minus sign, ${\cal Q}=-{\cal W}$. For $\phi=(n+1/2)\pi/2$, the system is topologically trivial, ${\cal Q}(k_{y})\equiv 0$, regardless of the choice of $\Delta_{\rm s},\Delta_{\rm t}$ (black dotted lines in Figs.\ \ref{fig:TopInv} and \ref{fig:Cond}). This can be understood as a consequence of spatial symmetry: For $\cos 2\phi=0$ the system fulfills 
\begin{equation}
H(k_x,k_y)=\sigma_y H(k_x,-k_y)\sigma_y\Rightarrow
r(k_y)=\sigma_y r(-k_y)\sigma_y.
\label{eq:spatialsym1}
\end{equation}
This is a symmetry condition of the type \eqref{rreflection}, with $a=y,b=0$, forcing the reflection matrix into the topologically trivial symmetry class CI (see Table \ref{tablereflect}). At $k_{y}=0$ the $\mathbb{Z}$ invariant ${\cal Q}$ vanishes, but the $\mathbb{Z}_{2}$ invariant ${\cal Q}_{0}$ can be nonzero. This happens for $\Delta_{\rm s}=0$, $\phi\neq 0$ (mod $\pi/2$), when ${\cal Q}_{0}=2$.

Fig.\ \ref{fig:Cond} shows how the topological invariant enforces the quantization of the angle-resolved conductance. First of all, $G/G_{0}=2$ whenever $|{\cal Q}|=2$ or ${\cal Q}_{0}=2$. For $\phi=0$ quantized plateaus at $G/G_{0}=1$ appear because of the spatial symmetry
\begin{equation}
r(k_y)=(\sigma_{y}\otimes\tau_{z}) r(-k_y)(\sigma_{y}\otimes\tau_{z}),
\label{eq:spatialsym2}
\end{equation}
which is a symmetry of the type (\ref{rreflection}) with $a,b=y,z$. This forces the reflection matrix into class BDI and ensures that the conductance is quantized for any nonzero ${\cal Q}$ (see Table \ref{tablereflect}).

\subsection{Anisotropic spin-orbit coupling}
\label{anisotropic}

\begin{figure}[tb]
\centerline{\includegraphics[width=1\linewidth]{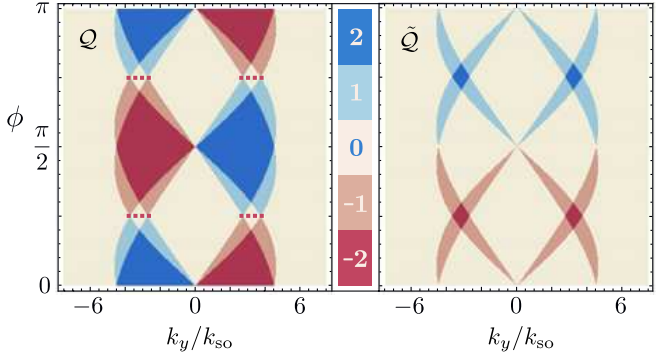}}
	\caption{Topological invariants ${\cal Q}$ (left panel) and $\tilde{\cal Q}$ (right panel) for an NS junction between a normal metal and the anisotropic Rashba superconductor of Sec.\ \ref{anisotropic}. The parameters chosen are: $\Delta_{\rm s}=E_{\rm so}$, $\Delta_{\rm t}=0$, $\mu=10\,E_{\rm so}$, $\mu_{\rm N}=30\,E_{\rm so}$. The $\mathbb{Z}_{2}$ invariant ${\cal Q}_{00}=2$ on the dotted red lines in the left panel.\\
	\label{fig:Aniso}}
\end{figure}

\begin{figure}[tb]
\centerline{\includegraphics[width=0.8\linewidth]{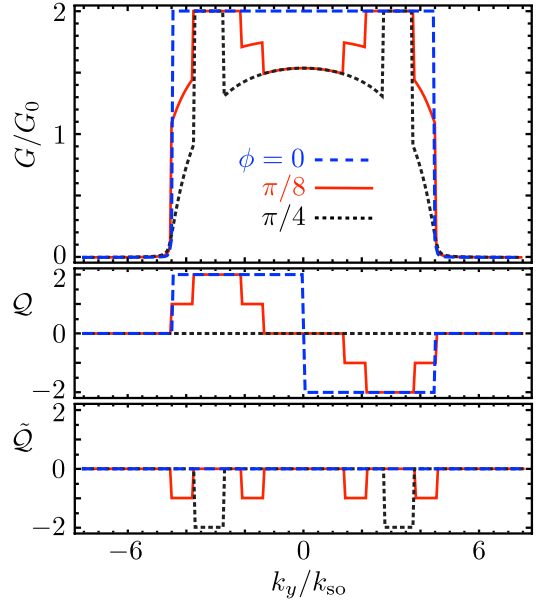}}
	\caption{Electrical conductance and $\mathbb{Z}$ topological invariants for three of the angles $\phi$ from Fig.\ \ref{fig:Aniso}. 
	\label{fig:Cond_aniso}}
\end{figure}

A strongly anisotropic dispersion, $m_x\gg m_y$, can produce an anisotropic spin-orbit coupling term of the form\cite{Ali10} $\bm{g}(\bm{k})=\lambda(0,-k_x,0)$. Topological invariants and conductance are plotted for the spin-singlet regime ($\Delta_{\rm t}=0$) in Figs.\ \ref{fig:Aniso} and \ref{fig:Cond_aniso}. There are two qualitative differences with the isotropic case of the previous subsections. 

First of all, for $\phi=n\pi/2$ the regions with $|{\cal Q}(k_y)|=1$ are missing. This can be explained by the spatial symmetry
\begin{equation}
r(k_y)=\tau_z r(-k_y) \tau_z,
\label{eq:spatialsym3}
\end{equation}
of the type (\ref{rreflection}) with $a,b=0,z$. As a consequence, see Table \ref{tablereflect}, the topological invariant ${\cal Q}(k_y)$ becomes a $2\mathbb{Z}$ invariant of class CII, excluding $|{\cal Q}(k_y)|=1$. 

Secondly, there is a unitary symmetry $\sigma_y r(k_y)\sigma_y=r(k_y)$ that holds for all $\phi$. This allows us to define an additional topological invariant,
\begin{equation}
\tilde{\cal Q}=\tfrac{1}{2}{\rm Tr}\,\sigma_{y}R=\tfrac{1}{2}{\rm Tr}\,\tau_x r, 
\end{equation}
following Sec.\ \ref{extrasym}. The topological invariants ${\cal Q}$ and $\tilde{\cal Q}$ are independent, in particular, $\tilde{\cal Q}(k_y)=\tilde{\cal Q}(-k_y)$ while ${\cal Q}(k_y)=-{\cal Q}(-k_y)$. Each topological invariant ${\cal Q}$ and $\tilde{\cal Q}$ gives a lower bound on the conductance. This explains the diamond-shaped regions in the phase diagram with a quantized conductance $G/G_{0}=2$, enforced by $|\tilde{\cal Q}|=2$.

There is a third invariant: At $\phi=(n+1/2)\pi/2$ the spatial symmetry $r(k_y)=r(-k_y)$ places the reflection matrix in symmetry class DIII. According to Eq.\ \eqref{QabDIII}, the corresponding $\mathbb{Z}_{2}$ invariant ${\cal Q}_{00}=2$ on the dotted red lines in the phase diagram.

This third invariant does not lead to additional constraints on the conductance, since we already have $\tilde{\cal Q}=2$ when ${\cal Q}_{00}=2$. But the two invariants ${\cal Q}$ and $\tilde{\cal Q}$ are both needed to explain the quantized conductance. The coexistence of two topological invariants is an unusual feature of this system.  

\section{Effects of angular averaging and disorder}
\label{disorder}

It may be possible to measure the angle-resolved conductance $G(\bm{k}_{\parallel})$,\cite{Mor99} but one typically measures the angular average. Moreover, disorder is detrimental for the conductance quantization if it mixes parallel momenta with different values of the topological invariant. In this section we investigate whether signatures of the conductance quantization can survive the effects of angular averaging and disorder.

We focus on the 2D Rashba superconductor of Sec.\ \ref{RashbaSC}, for $\Delta_{\rm t}=0$, $\phi=0$, when the topological invariant is given by Eq.\ \eqref{eq:TIHamiltonian2}. The angular average of the conductance for an interface of width $W$ is given by
\begin{equation}
G_{\rm NS}=\frac{W}{2\pi}\int_{-k_{\rm N}}^{k_{\rm N}}  dk_y\, G(k_y).\label{GNSdef}
\end{equation}
The reflection matrix, which determines $G(k_y)$ via Eq.\ \eqref{Gdef}, is calculated numerically using the square lattice discretization of the Hamiltonian \eqref{eq:Hamiltonian} (lattice constant $a=0.2\,l_{\rm so}$, $W=32\,l_{\rm so}$). Disorder is added to a strip $-L<x<0$ ($L= 31.6 \,l_{\rm so}$) of the normal region by means of a random on-site potential, distributed uniformly in $(-U_{0}/2,U_{0}/2)$. Results are averaged over 100 disorder realizations.

\begin{figure}[tb]
\centerline{\includegraphics[width=0.8\linewidth]{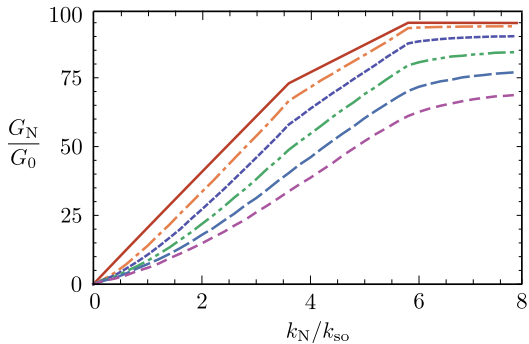}}
	\caption{Average conductance \eqref{GNSdef} of the NS junction as a function of the Fermi momentum $k_{\rm N}$ in the normal region, for various disorder strengths. The 2D Rashba superconductor has a $d_{xy}$-wave pair potential ($\phi=0$, $\Delta_{\rm t}=0$, $\Delta_{\rm s}=E_{\rm so}$, $\mu=10\,E_{\rm so}$). Disorder strengths from top to bottom curve: $U_0/E_{\rm so}=0,1,2,3,4,5$.
		 \label{fig:kfnscan}}
\end{figure}

In Fig.\ \ref{fig:kfnscan} we show the dependence of $G_{\rm NS}$ on the Fermi momentum $k_{\rm N}$ in the normal region. This is relevant if the normal region is a semiconductor, where one can vary $k_{\rm N}$ by a gate voltage. The quantization of $G(k_{y})$ manifests itself as a quantized slope of $G_{\rm NS}$ versus $k_{\rm N}$: the steep slope for $k_{\rm N}<k_{-}$ (where $|{\cal Q}|=2$) is reduced by a factor of two in the interval $k_{-}<k_{\rm N}<k_{+}$ (where $|{\cal Q}|=1$), and then is strongly suppressed for $k_{\rm N}>k_{+}$. This signature of the topological invariant gradually disappears with increasing disorder.

\begin{figure}[tb]
\centerline{\includegraphics[width=0.8\linewidth]{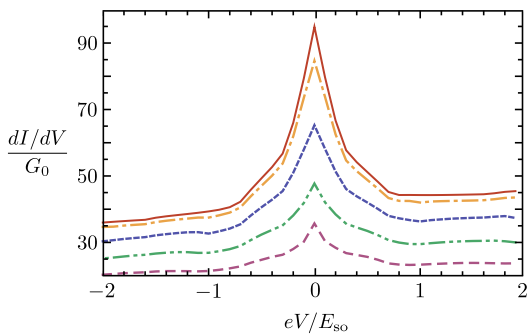}}
	\caption{Differential conductance of the NS junction for various disorder strengths. The parameters for the superconductor are the same as in Fig.\ \ref{fig:kfnscan}. In the normal region we have fixed $\mu_{\rm N}=25\,E_{\rm so}$. Disorder strengths from top to bottom curve: $U_0/E_{\rm so}=0,2.5,5,7.5,10$.
 \label{fig:energyscan}}
\end{figure}

Another signature can be seen for fixed $k_{\rm N}$ in the dependence of the differential conductance $dI/dV$ on the applied voltage $V$. As shown in Fig.\ \ref{fig:energyscan}, the peak in $dI/dV$ around $V=0$ is a superposition of two peaks with different widths, the narrower one originating from parallel momenta in the $|{\cal Q}|=2$ regions and the broader one from the $|{\cal Q}|=1$ regions. The single edge state of the latter regions couples more strongly to the continuum of the metal and thus has a larger width.

\section{Three-dimensional superconductors}
\label{sec:3D}

\subsection{Topological invariant for arc surface states}
\label{sec:arc} 

The topological invariants considered so far, and the resulting constraints on the angle-resolved conductance, apply both to 2D and 3D nodal superconductors. In this section we discuss features that are specific for 3D superconductors. The topological invariant ${\cal Q}(\bm{k}_{\parallel})$ of Sec.\ \ref{secQk} then counts dispersionless surface states, pinned to zero energy (the Fermi level) in a 2D region of parallel momentum $\bm{k}_{\parallel}=(k_1,k_2)$. The boundary of this flat band region is formed by nodal rings, closed contours of $\bm{k}_{\parallel}$ on which transmission through the superconductor is possible --- in other words, the superconducting gap vanishes for $\bm{k}=(k_\perp,\bm{k}_{\parallel})$.

The new feature that appears in a 3D superconductor is the possibility of zero-energy boundary states along a 1D arc connecting two nodal rings. Some aspects of their topological nature have been discussed in the Hamiltonian formulation of Ref.\ \onlinecite{Sch12}. Here we consider the alternative scattering formulation, and use it to obtain topological constraints on the conductance.

We consider a spatial symmetry on the 2D surface of a 3D superconductor, in which only one of the two components of parallel momentum is inverted:
\begin{equation}
r(k_1,k_2)=(\sigma_{a}\otimes\tau_{b})r(-k_1,k_2)(\sigma_{a}\otimes\tau_{b}).
\label{rsymmetryarc}
\end{equation}
Along the line $k_{2}=0$, this is a symmetry of the type (\ref{rreflection}), so we can follow Sec.\ \ref{restrictions} to introduce topological invariants ${\cal Q}_{ab}(k_{1})$. The resulting constraints on the angle-resolved conductance $G(k_{1},0)$ are summarized in Table \ref{tablereflect}. 

Alternatively, for $k_{1}=0$,  the symmetry (\ref{rsymmetryarc}) is of the type (\ref{rconservation}) with topological invariant $\tilde{\cal Q}(k_{2})$ from Eq.\ \eqref{QprimeAIIIdef}. The corresponding constraints on the conductance are discussed in Sec.\ \ref{extrasym}.

\subsection{Example}
\label{example:arc}

As an example, we apply these general considerations to the same Rashba Hamiltonian (\ref{eq:Hamiltonian}), but now with a 3D dispersion,
\begin{equation}
\epsilon(\bm{k})=(k_x^2+k_y^2+k_z^2)/2m-\mu.
\end{equation}
In the pair potential \eqref{Deltafdef} we set $f(\bm{k})\equiv 1$. This Hamiltonian applies to non-centrosymmetric s+p-wave superconductors of point group $C_{4\rm{v}}$. As described in Ref.\ \onlinecite{Sch12}, these superconductors have arc surface states connecting two nodal rings. They appear for example for the (011) surface orientation that we will consider in the following. The two components of parallel momentum on the surface are $k_1=k_x$ and $k_2= (k_y - k_z)/\sqrt{2}$. 

\begin{figure}[tb]
\centerline{\includegraphics[width=\linewidth]{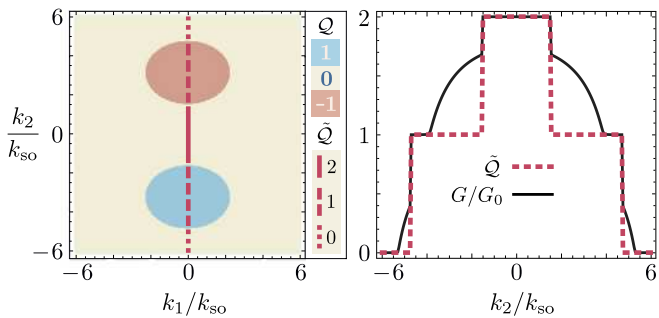}}
	\caption{Topological invariants and electrical conductance of an NS junction between a normal metal and an s+p-wave superconductor of point group $C_{4\rm{v}}$. The invariants ${\cal Q}$ and $\tilde{\cal Q}$ are plotted in the left panel. The right panel shows the conductance (black curve) and invariant $\tilde{\cal Q}$ (red dotted curve) along the line $k_1=0$.
	 Parameters chosen are $\Delta_{\rm s}=0.1\,E_{\rm so}$, $\Delta_{\rm t}=0.2\,E_{\rm so}$,
	$\mu=10\,E_{\rm so}$ and $\mu_{\rm N}=30\,E_{\rm so}$.
\label{fig:Inv_Cond_011}}
\end{figure}

We can obtain two topological invariants from the reflection matrix $r(k_1,k_2)$, plotted in the left panel of Fig.\ \ref{fig:Inv_Cond_011}. The first invariant
\begin{equation}
{\cal Q}(k_1,k_2)=\tfrac{1}{2}{\rm Tr}\,R(k_1,k_2)=\tfrac{1}{2}{\rm Tr}\,(\sigma_{y}\otimes\tau_{x})r(k_1,k_2)\label{Qdef3D}
\end{equation}
follows from chiral symmetry, see Sec.\ \ref{symmetry}, and is defined on the entire 2D plane of parallel momenta. This $\mathbb{Z}$ invariant is nonzero inside the regions bounded by the nodal rings, where it identifies a surface flat band. 

A second $\mathbb{Z}$ invariant appears as a consequence of the spatial symmetry
\begin{align}
&H(k_x,k_y,k_z)=(\sigma_x\otimes \tau_z ) H(-k_x,k_y,k_z)(\sigma_x\otimes \tau_z )\Rightarrow\nonumber\\
&\quad r(k_1,k_2)=(\sigma_x\otimes \tau_z ) r(-k_1,k_2)(\sigma_x\otimes \tau_z ).
\label{eq:arcsym}
\end{align}
The line $k_1=0$ connects the two nodal rings and on this line the invariant
\begin{equation}
\tilde{\cal Q}(k_2)=\tfrac{1}{2}{\rm Tr}\,(\sigma_x\otimes \tau_z )R(0,k_2)=-\tfrac{1}{2}{\rm Tr}\,(\sigma_z\otimes \tau_y )r(0,k_2) \label{tildeQdef3D}
\end{equation}
can take on a nonzero value.

The non-trivial invariants enforce a lower bound on the conductance, as is illustrated in the right panel of Fig. \ref{fig:Inv_Cond_011}. This leads to a quantized conductance $G/G_0=2$ along the line $k_1=0$.

The symmetry (\ref{eq:arcsym}) produces arc surface states on all surfaces parallel to the $x$-direction. For the (010) surface analyzed in Ref.\ \onlinecite{Sch12} there is an additional spatial symmetry, $r(k_x,k_z)=r(k_x,-k_z)$. For $k_x=0$ this additional symmetry allows for the $\mathbb{Z}_2$ invariant ${\cal Q}_{00}=1+\text{Pf}\,\sigma_y r(0,k_z)$, in addition to the $\mathbb{Z}$ invariant \eqref{tildeQdef3D}. For other surface orientations $(0nm)$ only the $\mathbb{Z}$ invariant is responsible for the arc states.

\section{Conclusion}
\label{conclusion}

In conclusion, we have constructed a topological invariant ${\cal Q}(\bm{k}_{\parallel})$ of the Andreev reflection matrix at the interface between a time-reversal symmetric nodal superconductor and a normal metal. In the absence of a tunnel barrier, this interface has no zero-energy boundary states, but the topologically nontrivial phase can still be detected in the angle-resolved conductance $G(\bm{k}_{\parallel})$. A variety of symmetry classes can be realized (AIII, BDI, CI, CII, DIII), by allowing for additional unitary symmetries. The corresponding topological invariants are given by a trace or Pfaffian of the reflection matrix. 

Many of these topological invariants have been studied before in the Hamiltonian formulation for an infinite system.\cite{Sat11,Sch11,Sat06,Tan09,Bry11,Sch12} The scattering formulation presented here makes it possible to directly relate ${\cal Q}(\bm{k}_{\parallel})$ to $G(\bm{k}_{\parallel})$. We have systematically examined when a nontrivial topological invariant enforces a quantized conductance, and when it only provides a lower bound. This approach can identify surface flat bands (within nodal rings) as well as arc states (connecting nodal rings), even when these zero-energy boundary states have merged with the continuum of states in the normal metal.

We have applied the general theory to 2D and 3D superconductors with spin-singlet and spin-triplet pairing mixed by Rashba spin-orbit coupling. The appearance of a quantized conductance has allowed us to verify known topological invariants and to identify new ones. In particular, in the 2D case of a strongly anisotropic spin-orbit coupling, we have shown the coexistence of two topological invariants --- which provide independent constraints on the conductance.

To make contact with experiments, the effects of angular averaging and impurity scattering on the conductance quantization have been investigated by numerical simulation of a disordered NS interface.

\acknowledgments
We thank A. R. Akhmerov and M. Wimmer for valuable discussions. Numerical simulations were performed with the {\sc kwant} software package, developed by A. R. Akhmerov, C. W. Groth, X. Waintal, and M. Wimmer. Our research was supported by the Dutch Science Foundation NWO/FOM, by an ERC Advanced Investigator Grant, by the EU Project IP-SOLID, and by the EU network NanoCTM.

\appendix

\section{Topological invariant counts number of unit Andreev reflection eigenvalues}
\label{unitrhonproof}

\subsection{Proof for the $\bm{\mathbb{Z}}$ invariant}
\label{proofZ}

The Hermitian matrix $R_{eh}^{2}$ has eigenvalues $\rho_{n}\in[0,1]$. We wish to prove that at least $|{\cal Q}|$ of these Andreev reflection eigenvalues are equal to unity. 

Let $\phi$ be an eigenvector of $R_{eh}$ with eigenvalue $\lambda$. Assume $\lambda\neq\pm 1$ (so $\rho=\lambda^{2}<1$). Since $R_{eh}^{2}=1-R_{hh}^{\dagger}R_{hh}$, the vector $\phi'=R_{hh}\phi$ cannot vanish. Since $R_{he}R_{hh}=-R_{hh}R_{eh}$, it then follows that $\phi'$ is an eigenvector of $R_{he}$ with eigenvalue $\mu=-\lambda$.

Now consider the $\mathbb{Z}$ topological invariant ${\cal Q}=\frac{1}{2}\,\sum_{n}(\lambda_{n}+\mu_{n})$ in symmetry class AIII. The eigenvalues $\lambda_{n}\neq\pm 1$ of $R_{eh}$ are cancelled by an eigenvalue $\mu_{n}=-\lambda_{n}$ of $R_{he}$. The cancellation can only be avoided for the $M$ eigenvalues $\lambda_{n}$ equal to $\pm 1$, resulting in $|{\cal Q}|\leq M$ --- as we set out to prove.

\subsection{Proof for the $\bm{\mathbb{Z}_{2}}$ invariant}
\label{proofZ2}

For any $4\times 4$ antisymmetric matrix $A$ with a block structure,
\begin{equation}
A=\begin{pmatrix}
A_{11}&A_{12}\\
A_{21}&A_{22}
\end{pmatrix}=-A^{\rm T},\label{Adef}
\end{equation}
the Pfaffian is given by
\begin{equation}
{\rm Pf}\,A=-{\rm Det}\,A_{12}-\tfrac{1}{2}\,{\rm Tr}\,A_{11}A_{22}.
\end{equation}

We apply this identity to the antisymmetric matrix $\sigma_{y}r$ at ${\bm k}_{\parallel}=0$, to obtain the $\mathbb{Z}_{2}$ topological invariant in symmetry class DIII,
\begin{align}
{\cal Q}_{0}&=1-{\rm Det}\,R_{eh}-\tfrac{1}{2}\,{\rm Tr}\,R_{ee}R_{hh}\nonumber\\
&=1-{\rm Det}\,R_{eh}-\tfrac{1}{2}\,{\rm Tr}\,(1-R_{eh}^{2}).\label{Q0Rresult}
\end{align}
In terms of the two eigenvalues $\lambda_{1},\lambda_{2}\in[-1,1]$ of $R_{eh}$ this reduces to
\begin{equation}
{\cal Q}_{0}=\tfrac{1}{2}(\lambda_{1}-\lambda_{2})^{2}.\label{Q0lambda}
\end{equation}

Since by construction ${\cal Q}_{0}$ equals either $0$ or $2$, we have either ${\cal Q}_{0}=0\Leftrightarrow\lambda_{1}=\lambda_{2}$ or ${\cal Q}_{0}=2\Leftrightarrow\lambda_{1}=-\lambda_{2}=\pm 1$. This shows that at least ${\cal Q}_{0}$ of the Andreev reflection eigenvalues $\rho_{n}=\lambda_{n}^{2}$ are equal to unity.

\section{Proof of Eq.\ \eqref{GtildeQconstraint}}
\label{app:condconstr3}

We consider the topological invariant \eqref{QprimeAIIIdef}, constructed from the matrix $\tilde{R}=(\sigma_{a}\otimes\tau_{b})R$ with $b\in\{x,y\}$, and wish to proof the constraint \eqref{GtildeQconstraint} on the conductance. This amounts to a proof that at least $|\tilde{\cal Q}|$ of the Andreev reflection eigenvalues are equal to zero. 

We define the Hermitian matrix 
\begin{align}
\bar{R}=\begin{pmatrix} \bar{R}_{{ee}} &  \bar{R}_{{eh}}\\  \bar{R}_{{he}}&  \bar{R}_{{hh}} \end{pmatrix}\equiv \begin{cases}
\tilde{R}(\bm{k}_{\parallel})&{\rm if}\;\;\tilde{R}^{2}=1,\\
i\tilde{R}(\bm{k}_{\parallel})&{\rm if}\;\;\tilde{R}^{2}=-1.
\end{cases}
\end{align}
Let $\phi$ be an eigenvector of $\bar{R}_{ee}$ with eigenvalue $\lambda$. Assume $\lambda\neq\pm 1$. Since $\bar{R}_{ee}^{2}=1-\bar{R}_{he}^{\dagger}\bar{R}_{he}$, the vector $\phi'=\bar{R}_{he}\phi$ cannot vanish. With $\bar{R}_{he}\bar{R}_{ee}=-\bar{R}_{hh}\bar{R}_{he}$, it then follows that $\phi'$ is an eigenvector of $\bar{R}_{hh}$ with eigenvalue $\mu=-\lambda$.

Now since $\tilde{{\cal Q}}=\frac{1}{2}\,\text{Tr}(\bar{R}_{ee}+\bar{R}_{hh})=\frac{1}{2}\,\sum_{n}(\lambda_{n}+\mu_{n})$, the eigenvalues $\lambda_{n}\neq\pm 1$ of $R_{ee}$ are cancelled by eigenvalues $\mu_{n}=-\lambda_{n}$ of $R_{hh}$ in the expression for the topological invariant. The cancellation can only be avoided for the $M$ eigenvalues $\lambda_{n}$ equal to $\pm 1$, resulting in $|\tilde{\cal Q}|\leq M$. The existence of at least $|\tilde{\cal Q}|$ unit eigenvalues of $\bar{R}^\dag_{ee}\bar{R}_{ee}=\bar{R}_{ee}^2$ is equivalent to the existence of at least $|\tilde{\cal Q}|$ zero Andreev reflection eigenvalues and thereby proves Eq.\ \eqref{GtildeQconstraint}.

\section{Equality of conductance and topological invariant in class BDI}
\label{Berideg}

A topologically nontrivial $4\times 4$ reflection matrix in class BDI has either $|{\cal Q}_{ab}|=2$ or $|{\cal Q}_{ab}|=1$. In the former case the inequality \eqref{GQab} is saturated, because $G/G_{0}\leq 2$, but in the latter case it provides only a lower bound on the conductance. We now wish to show that the inequality can be sharpened to an equality for three of the six spatial symmetries \eqref{rreflection} in class BDI. More precisely, we will show that $|{\cal Q}_{ab}|=1$ implies $G/G_{0}=1$ for $(a,b)\in\{(y,z),(x,0),(z,0)\}$.

For each of these three cases the symmetry relation \eqref{calTsymmetry} implies that $R_{he}=\sigma_{a}R_{eh}^{\rm T}\sigma_{a}$, so ${\rm Tr}\,R_{he}={\rm Tr}\,R_{eh}$. Denote the eigenvalues of $R_{eh}$ and $R_{he}$ by $\lambda_{1},\lambda_{2}$ and $\mu_{1},\mu_{2}$, respectively. (All are real numbers in the interval $[-1,1]$.) The equality of the traces gives $\lambda_{1}+\lambda_{2}=\mu_{1}+\mu_{2}$. The topological invariant \eqref{QabBDI} determines the sum $\lambda_{1}+\lambda_{2}+\mu_{1}+\mu_{2}=2{\cal Q}_{ab}$, hence $\lambda_{1}+\lambda_{2}={\cal Q}_{ab}$.

Because classes BDI and AIII have the same expression for the topological invariant, we may apply the result of App.\ \ref{proofZ} that at least $|{\cal Q}_{ab}|$ of the $\lambda_{n}$'s equal $\pm 1$. If we take $|{\cal Q}_{ab}|=1$, $|\lambda_{1}|=1$, then necessarily $\lambda_{2}=0$. The dimensionless conductance $G/G_{0}=\lambda_{1}^{2}+\lambda_{2}^{2}$ thus equals unity, as we set out to prove.

Our finding can be seen in a broader context as a manifestation of B\'{e}ri degeneracy of Andreev reflection eigenvalues:\cite{Ber09} The charge-conjugation symmetry \eqref{calPHsymmetry}, with $(a,b)\in\{(y,z),(x,0),(z,0)\}$, enforces a twofold degeneracy of the Andreev reflection eigenvalues $\rho_{n}=\lambda_{n}^{2}$ that can only be avoided if $\rho_{n}$ equals $0$ or $1$.


\begin{thebibliography}{99}
\bibitem{note1} The topological invariant ${\cal Q}$ is restricted to the integers $0,\pm 1,\pm 2$ because we consider only spin and electron-hole degrees of freedom. Further integer values become possible if multiple valleys or layers produce additional internal degrees of freedom, which is why ${\cal Q}$ is a called a $\mathbb{Z}$ invariant. This does not apply to ${\cal Q}_{0}$, which is restricted to $0,2$ irrespective of additional degrees of freedom, hence the name $\mathbb{Z}_{2}$ invariant.
\bibitem{note2} To see that ${\cal Q}_{0x}=1+{\rm Pf}\,R$ is independent of $G$, note that the transformation $R\mapsto ORO$ with $O={\rm diag}\,(1,1,1,-1)$ switches the sign of the Pfaffian --- while leaving the conductance unchanged. In a similar way one can show that also the DIII invariants ${\cal Q}_{xy}$ and ${\cal Q}_{zy}$ are independent of $G$.
\end{thebibliography}
\end{document}